\documentclass[a4paper,12pt,nonacm]{acmart}
\usepackage{graphicx}
\usepackage{tikz}
\usepackage{geometry}
\usepackage{xcolor}
\usepackage[T1]{fontenc}
\usepackage{tgbonum}
\usepackage{lmodern}
\usepackage{tgtermes}
\usepackage{tgpagella}
\usepackage{tgschola}
\usepackage{mathptmx}
\usepackage{utopia}
\usepackage{palatino}
\usepackage{bookman}
\usepackage{charter}
\usepackage{tabularx}
\usepackage{longtable}
\usepackage{hyperref}
\usepackage{titlesec}
\usepackage{fontawesome}
\usepackage{xcolor}
    
\definecolor{offwhite}{RGB}{255,255,255}

\newcommand{\papertitle}{Tracking the 2024 US Presidential Election Chatter on Tiktok: A Public Multimodal Dataset}
\newcommand{\paperauthors}{Gabriela Pinto, Charles  Bickham, Tanishq Salkar, Luca Luceri, Emilio Ferrara}
\newcommand{\paperaffiliation}{University of Southern California} 

\titlespacing*{\subsection}
{0pt}{*0.5}{*0.25} 
\newcommand{\blueheart}{\textcolor{blue}{\faHeart}}

\begin{document}

\begin{titlepage}
    \begin{tikzpicture}[remember picture, overlay]
        \node[anchor=north west, inner sep=0] at (current page.north west) {
            \includegraphics[width=\paperwidth, height=\paperheight, trim=275 375 275 375]{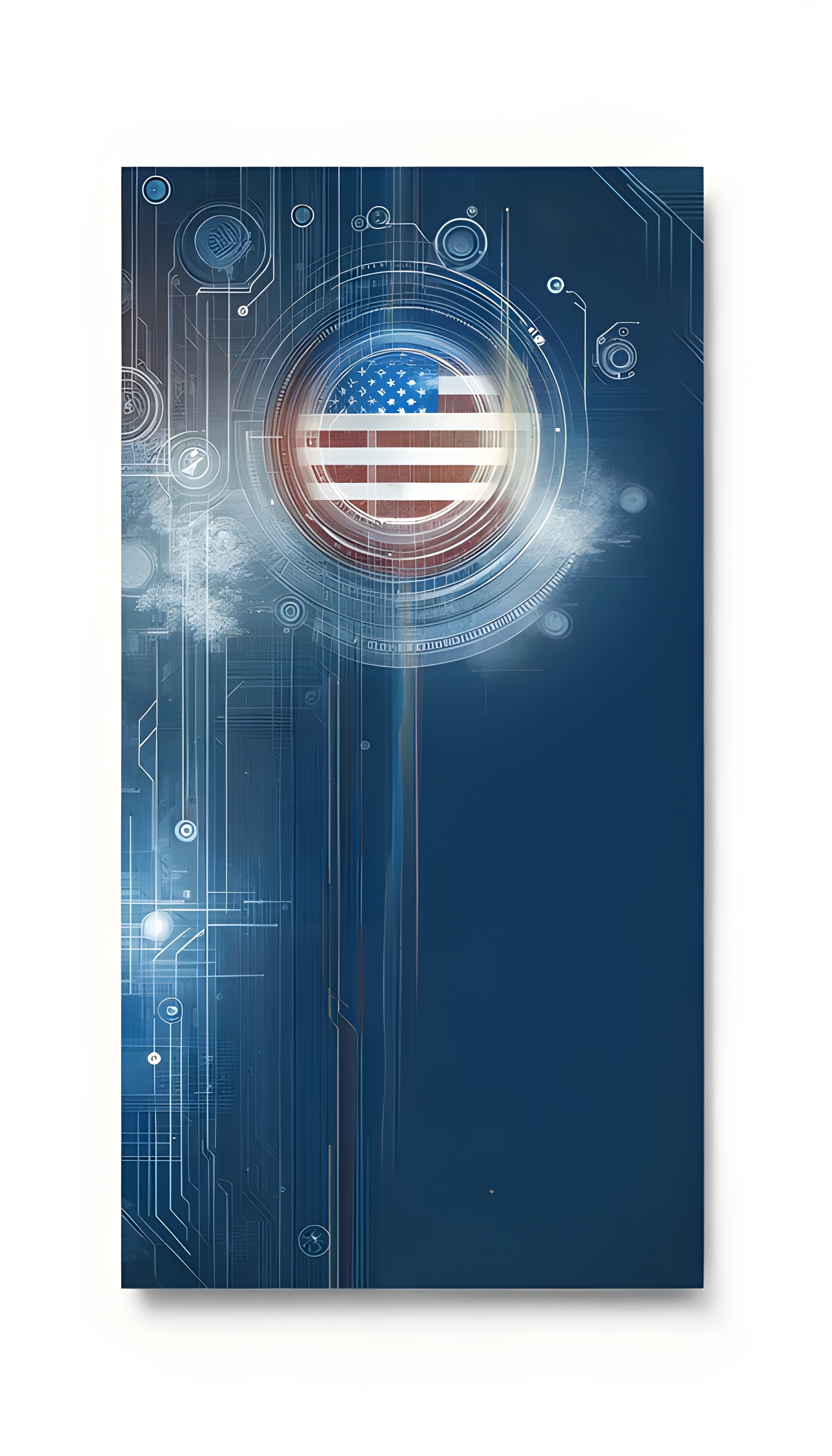}
        };
        \node[anchor=center, yshift=-9cm] at (current page.center) {
            \begin{minipage}{\textwidth}
                \raggedleft
                \color{offwhite}
                {\Huge \bfseries \fontfamily{qtm}\selectfont The 2024 Election Integrity Initiative }
                
                \vspace{1.5cm}
                
                {\LARGE \fontfamily{qtm}\selectfont \papertitle}
                
                \vspace{1.5cm}
                
                {\large \fontfamily{qtm}\selectfont \paperauthors}
                
                \vspace{1cm}
                
                {\Large \fontfamily{qtm}\selectfont \paperaffiliation}
                
                \vfill
                
                {\Large \fontfamily{qtm}\selectfont HUMANS Lab -- Working Paper No. 2024.3}
            \end{minipage}
        };
    \end{tikzpicture}
\end{titlepage}

\noindent{\LARGE \fontfamily{qtm}\selectfont \papertitle}

\vspace{0.5cm}

\noindent{\large \fontfamily{qtm}\selectfont \paperauthors}

\noindent{\large \fontfamily{qtm}\selectfont \textit{\paperaffiliation}}

\section*{Abstract}
This paper documents our release of a large-scale data collection of TikTok posts related to the upcoming 2024 U.S. Presidential Election. Our current data comprises 1.8 million videos published between November 1, 2023, and May 26, 2024. Its exploratory analysis identifies the most common keywords, hashtags, and bigrams in both Spanish and English posts, focusing on the election and the two main Presidential candidates, President Joe Biden and Donald Trump. 

We utilized the TikTok Research API, incorporating various election-related keywords and hashtags, to capture the full scope of relevant content. To address the limitations of the TikTok Research API, we also employed third-party scrapers to expand our dataset. The dataset is publicly available at \href{https://github.com/gabbypinto/US2024PresElectionTikToks}{https://github.com/gabbypinto/US2024PresElectionTikToks}.

\section*{Introduction}
Social media has profoundly transformed electoral politics, emerging as a critical platform for disseminating election-related information. This shift has garnered significant attention from researchers \cite{jungherr2016twitter, kratzke2017btw17, deb2019perils, luceri2019evolution, abilov2021voterfraud2020}. Twitter, in particular, has been instrumental in providing datasets that aid the study of global geopolitical events \cite{davis2016osome, chen2022election2020, chen2023tweets}. 

Meanwhile, TikTok, a short-form video app, has rapidly grown into a major platform for engaging and informing users on a variety of topics. With over a billion users worldwide,\footnote{\url{https://whatsthebigdata.com/tiktok-statistics/}} TikTok is especially popular among adolescents \cite{montag2021psychology}. Given its rising popularity, TikTok has the potential to become a central hub for disseminating information about the upcoming 2024 election.

As TikTok's influence expands, it has become a strategic platform for politicians aiming to reach young voters. For instance, Donald Trump joined TikTok in June 2024 and quickly amassed over 6 million followers,\footnote{\url{https://cnn.com/2024/06/02/politics/donald-trump-joins-tiktok/index.html}} while Joe Biden, with more than 373,000 followers, has posted over 200 videos since February 2024.\footnote{\url{https://cnn.com/2024/06/12/tech/tiktok-pew-research-politics-x/index.html}} The platform's growing popularity and diverse content formats provide a wealth of data that can reflect electoral sentiments and trends.

In this paper, we introduce the TikTok 2024 U.S. Presidential election dataset. This multimodal dataset, which includes both video and text data, aims to offer researchers a comprehensive view of TikTok's role in the election. By capturing political discourse on the platform and updating keywords as political campaigns progress, we hope this dataset will enable valuable insights into the evolving landscape of political communication and engagement on TikTok.

\section*{Method of Data Collection}
\subsection*{TikTok API}
The TikTok Research API\footnote{\url{https://developers.tiktok.com/doc/research-api-specs-query-videos/}} facilitates the retrieval of detailed information regarding accounts and content on TikTok.  
Account-related data like user profiles, followers and following lists, liked videos, pinned videos, and reposted videos.
Content details include comments, captions, subtitles, and the number of comments, shares, and likes that a video receives.
Accessing data via this API requires an access token obtained using the user's credentials (i.e., client key and secret key). This bearer token, accompanied by specific parameters in the request body, allows the fetch of desired data.
\\
\subsection*{Query}
The TikTok API allows us to specify specific parameters to get desired data from the API. We define the parameters in the request body, the authentication token, and the format of the response expected from the API. These parameters include the range of dates of publication and the metadata fields of the videos needed in the response. The API allows us to get the desired data of videos published in specific geographical regions, videos made on particular topics, videos using different languages, and responses to the videos in the form of likes, shares, and comments.
\begin{itemize}
    \item Start date: The starting date and time window from which to extract data.
    \item End date: The ending date and time window until which to extract data.
    \item Fields: The specific metadata of the videos required in the response, such as video identifier (id), description (video\_description), creation time (create\_time), region code (region\_code), share count (share\_count), view count (view\_count), like count (like\_count), comment count (comment\_count), music identifier (music\_id), hashtag names (hashtag\_names), and username (username).
    \item Max\_count: Specifies the number of records to be returned in a call, with a maximum of 100 records per request.
    
    \item Query Body: The TikTok API also allows the use of a query body to create more detailed queries, similar to SQL, by using logical operators such as AND and OR on specified fields like region code, hashtag names, likes etc. For example, to collect metadata of videos published in the United States with the hashtag "\#elections2024" by specifying "US" in region\_code field and \#elections2024 in hashtag field, the necessary conditions can be specified in the query body.
\end{itemize}

In this study, we are collecting metadata of videos published in the US using the TikTok Research API. The data collection encompasses all the fields mentioned previously, including share count and view count, such as count, comment count, music identifier, hashtag names, username, etc. Additionally, we are incorporating a list of keywords and hashtags, as detailed in the subsequent sections.

\subsection*{Keywords and Hashtags}

The hashtags/keywords detailed in Table \ref{tab:words_hashtags} (cf. Appendix) illustrate the inclusion and exclusion of keywords within the indicated phases. Table \ref{tab:phases} details each phase's start and end dates. Those marked with ‘-’ means that their corresponding keywords/hashtags were excluded from the query; conversely, those marked with ‘+” represent the inclusion. We created the list of keywords and phases based on significant political events that gained massive media attention.

We set the end of \textit{Phase 1} on January 15, 2024, the day of the Iowa caucus.\footnote{\href{https://www.nytimes.com/interactive/2024/01/15/us/elections/results-iowa-caucus.html}{https://www.nytimes.com/interactive/2024/01/15/us/elections/results-iowa-caucus.html}} Thus, January 16, 2024, was the new start date for \textit{Phase 2}. In \textit{Phase 2} and for subsequent phases, we analyzed the most common keywords/hashtags within our data at the time of collection. The most frequent keywords and hashtags related to the US Presidential Elections were added to the query. On January 21, 2024, Republican nominee Ron DeSantis suspended his presidential campaign.\footnote{\href{https://www.politico.com/news/2024/01/21/desantis-ends-presidential-campaign-00136839}{https://www.politico.com/news/2024/01/21/desantis-ends-presidential-campaign-00136839}
} Consequently, we removed Ron DeSantis in \textit{Phase 2}.  On February 7, 2024, Democratic nominee Marianne Williamson announced the end of her presidential campaign due to her loss at the Nevada Democratic Primary to President Joe Biden.\footnote{\href{https://www.politico.com/news/2024/02/07/marianne-williamson-drops-out-2024-00140297}{https://www.politico.com/news/2024/02/07/marianne-williamson-drops-out-2024-00140297}} Therefore, February 8, 2024, became the start date for \textit{Phase 3}, where keywords/hashtags related to Marianne Williamson and Ron DeSantis were removed. \textit{Phase 3} spanned from February 8, 2024, to February 27, 2024, since we wanted to reevaluate our data and extract the most popular keywords and hashtags. We updated the query by including the most reoccurring and relevant keywords and hashtags. \textit{Phase 4} spanned between February 28, 2024 to March 5, 2024. The end date of March 5, 2024 was established due to the announcement of Democratic candidate Dean Phillips\footnote{\href{https://www.politico.com/news/2024/03/06/dean-phillips-drops-out-00145403}{https://www.politico.com/news/2024/03/06/dean-phillips-drops-out-00145403}} and Republican Nikki Haley\footnote{\href{https://www.wsj.com/politics/elections/nikki-haley-drops-out-2024-presidential-election-625277ca}{https://www.wsj.com/politics/elections/nikki-haley-drops-out-2024-presidential-election-625277ca}} announced their end to their presidential campaign. Given the lack of media attention on social media on Dean Phillips, we removed the related keywords/hashtags for \textit{Phase 5}. March 6, 2024, was the start date of \textit{Phase 5}; however, keywords and hashtags related to Nikki Haley were included until June 17, 2024, since she received more media attention than Dean Phillips. 

\begin{table}[t]
  \centering  
  \caption{Phases and the dates ranges.}
  \begin{tabularx}{0.8\textwidth}{|X|X|}
    \hline
    Phase \# & (MM/DD/YYYY)-(MM/DD/YYYY)\\
    \hline
    1 &  11/01/2023 - 01/15/2024\\
    \hline
    2 &  01/16/2024 - 02/07/2024\\
    \hline
    3 &  02/08/2024 - 02/27/2024\\    
    \hline
    4 &  02/28/2024 - 03/05/2024\\    
    \hline
    5 &  03/06/2024 - 05/26/2024\\    
    \hline
  \end{tabularx}
  \label{tab:phases}
\end{table}

\section*{Exploratory Analysis}
\subsection*{Data Access}
The dataset is publicly available on Github,\footnote{\href{https://github.com/gabbypinto/US2024PresElectionTikToks}{https://github.com/gabbypinto/US2024PresElectionTikToks}} where we provide the ID of each video collected. We will update the repository consistently for future collections as we collect more videos. In compliance with the Terms of Service of TikTok's Research API, we can only publish the IDs since publishing data that risks users' privacy is strictly prohibited.

\subsection*{Summary of Data}
In Table~\ref{tab:sumStats}, we present a summary of the statistics in the current version of the dataset presented in this paper. The proportion of the number of transcripts to the total number of videos is approximately 14.7\% collected. The low proportion of TikTok-generated transcripts is one of the limitations of the TikTok Research API. Therefore, to provide more rich linguistic data on the published content, we will upload transcripts generated by Whisper\footnote{\href{https://github.com/openai/whisper}{https://github.com/openai/whisper}} in our GitHub repository.  

\begin{table}[t]
\caption{Summary statistics of the dataset.}
\begin{tabular}{|l|l|}
    \hline
    Number  of videos & 1,799,333 \\ \hline  
    Number  of transcripts & 266,202 \\ \hline 
    Number  of comments & 93,776,960 \\ \hline
    Number  of views   & 3,535,119,560   \\ \hline
    Number  of likes   & 871,474,292   \\ \hline
    Number  of shares   & 140,038,704   \\ \hline
    Number  of unique users & 432,951 \\ \hline
\end{tabular}
\label{tab:sumStats}
\end{table}

\subsection*{Number of Videos over time}
\begin{figure}[t]
\centering
\includegraphics[clip, trim=5 3 0 20, height=0.3\textheight]{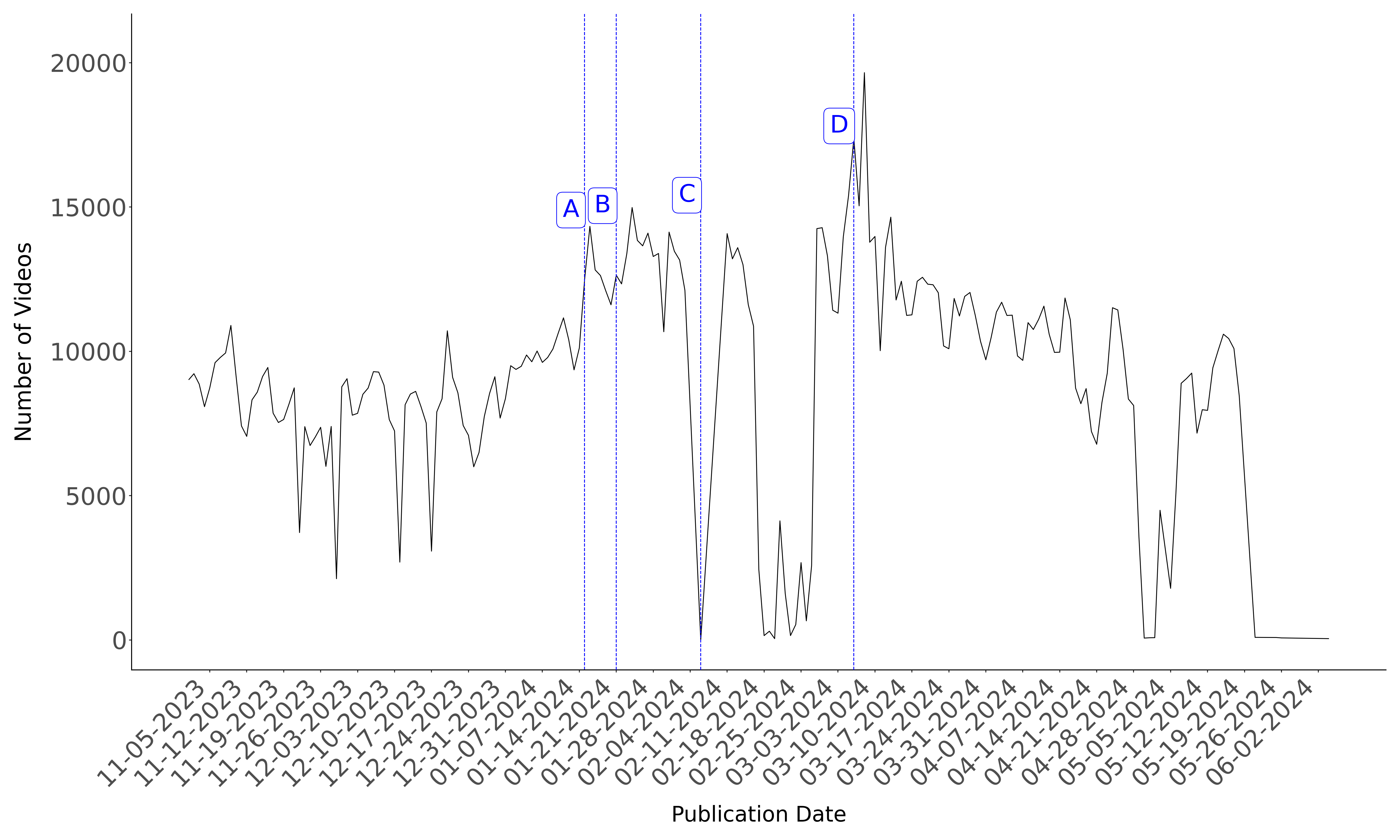}
\centering \footnotesize
\begin{tabularx} {0.7\textwidth}{|l|l|X|} %
\hline
\textbf{\#} & \textbf{Date} & \textbf{Description} \\ 
\hline

A & 01-15-2024 & Trump wins the Iowa Caucus \\ 
\hline

B & 01-21-2024 & Ron DeSantis suspends his campaign  \\ 
\hline

C & 02-06-2024 & Marianne Williamson suspends her campaign \\ 
\hline

D & 03-06-2024 & Dean Phillips and Nikki Haley suspend their campaigns \\ 
\hline

\end{tabularx}
\caption{Timeline of events and volume of TikTok posts.}
\label{fig:videostime}
\end{figure}

In this dataset's first deployment, we collected video metadata posted between November 1, 2023, and May 26, 2024. Figure~\ref{fig:videostime} the number of videos collected concerning its publication date. It is important to note that there are notable gaps within our data, which is content that we couldn't collect with respect to its publication date via the TikTok Research API. Table~\ref{tab:gaps} contains each gap's start and end date inclusively. To solve this issue, we currently use third-party scrapers to collect video content published within the stated dates. 

In addition, we also provided the dates of critical political events during the election cycle. We are currently using additional scrapers\footnote{\href{https://github.com/davidteather/TikTok-Api}{https://github.com/davidteather/TikTok-Api}} to collect more data and fill in the gaps to obtain the complete discourse on TikTok from November 1, 2023, to January 1, 2025. 

\setlength{\belowcaptionskip}{-0.7cm}

\begin{table}[t]
\footnotesize
\caption{The range for each gap within our dataset.}
\begin{tabular}{|l|l|}
    \hline
    Start Date & End Date \\ \hline  
    02-04-2024 & 02-10-2024 \\ \hline 
    05-04-2024 & 05-04-2024 \\ \hline 
    05-19-2024 & 05-20-2024 \\ \hline 
    05-22-2024 & 05-24-2024   \\ \hline 
\end{tabular}
\label{tab:gaps}
\end{table}

\subsection*{Top Keywords mentioned in the video descriptions}
\begin{figure}[t]
  \centering
\includegraphics[height=0.3\textheight]{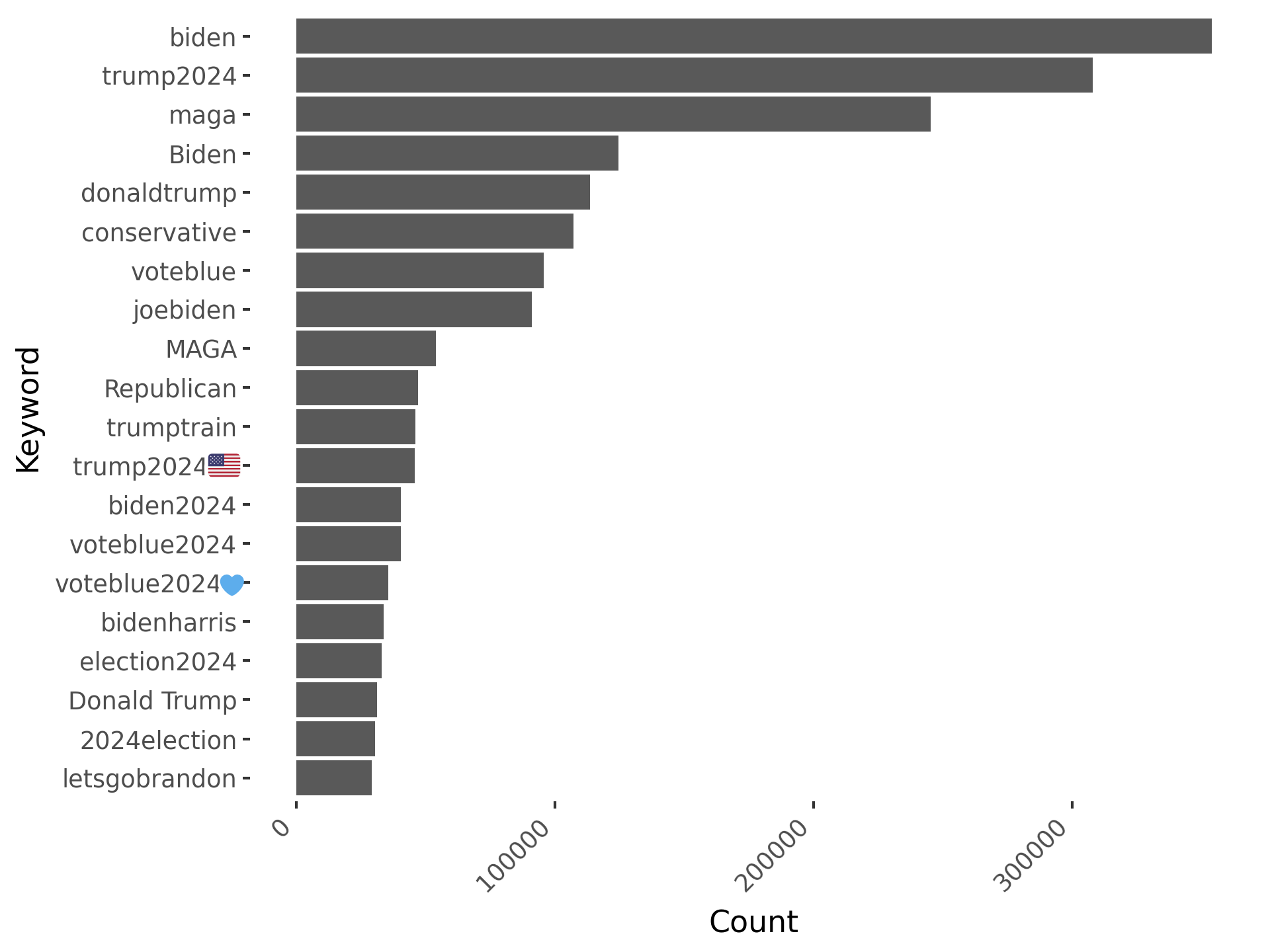}
  \caption{Top Keywords within our query that appear in the `video\_description' attribute}
  \label{fig:top_phrases}
\end{figure}

The intersection between the words in the 'video\_descriptions' attribute (the caption the user has posted along with the video) and the keywords shown in Table~\ref{tab:words_hashtags}. Figure~\ref{fig:top_phrases}, shows the 20 most reoccurring keywords and their count. The frequency reflects that the video content is mainly related to President Joe Biden and Donald Trump.

\subsection*{Top Hashtags}
Table~\ref{tab:tophastags} presents the 20 most frequent hashtags for each video. Each video collected through the TikTok Research API provides a list of hashtags labeled 'hashtag\_names.' Overall, the hashtags reflect that the videos labeled by the creator were related to Donald Trump, President Joe Biden, and the right-leaning ideology through keywords such as "maga" ("Make America Great Again") and "republican."

\begin{table}[t]
\centering\footnotesize
\caption{Top 20 Most Frequently Occurred Hashtags.}
\begin{tabular}{ll}
\toprule
\textbf{Tag} & \textbf{Count} \\
\midrule
trump        & 407,541 \\
trump2024    & 285,840 \\
biden        & 198,247 \\
maga         & 170,354 \\
duet         & 146,676 \\
republican   & 144,067 \\
usa          & 130,607 \\
donaldtrump  & 113,514 \\
politics     & 107,345  \\
democrat     & 99,747  \\
news         & 95,608  \\
america      & 92,740  \\
joebiden     & 86,267  \\
trending     & 85,231  \\
conservative & 76,203  \\
fjb          & 69,401  \\
capcut       & 62,360  \\
voteblue     & 61,309  \\
democrats    & 53,799  \\
election     & 52,119  \\
\bottomrule
\end{tabular}
\label{tab:tophastags}
\end{table}

\subsection*{Language Detection}
\begin{figure}[t]
  \centering
\includegraphics[height=0.55\textheight]{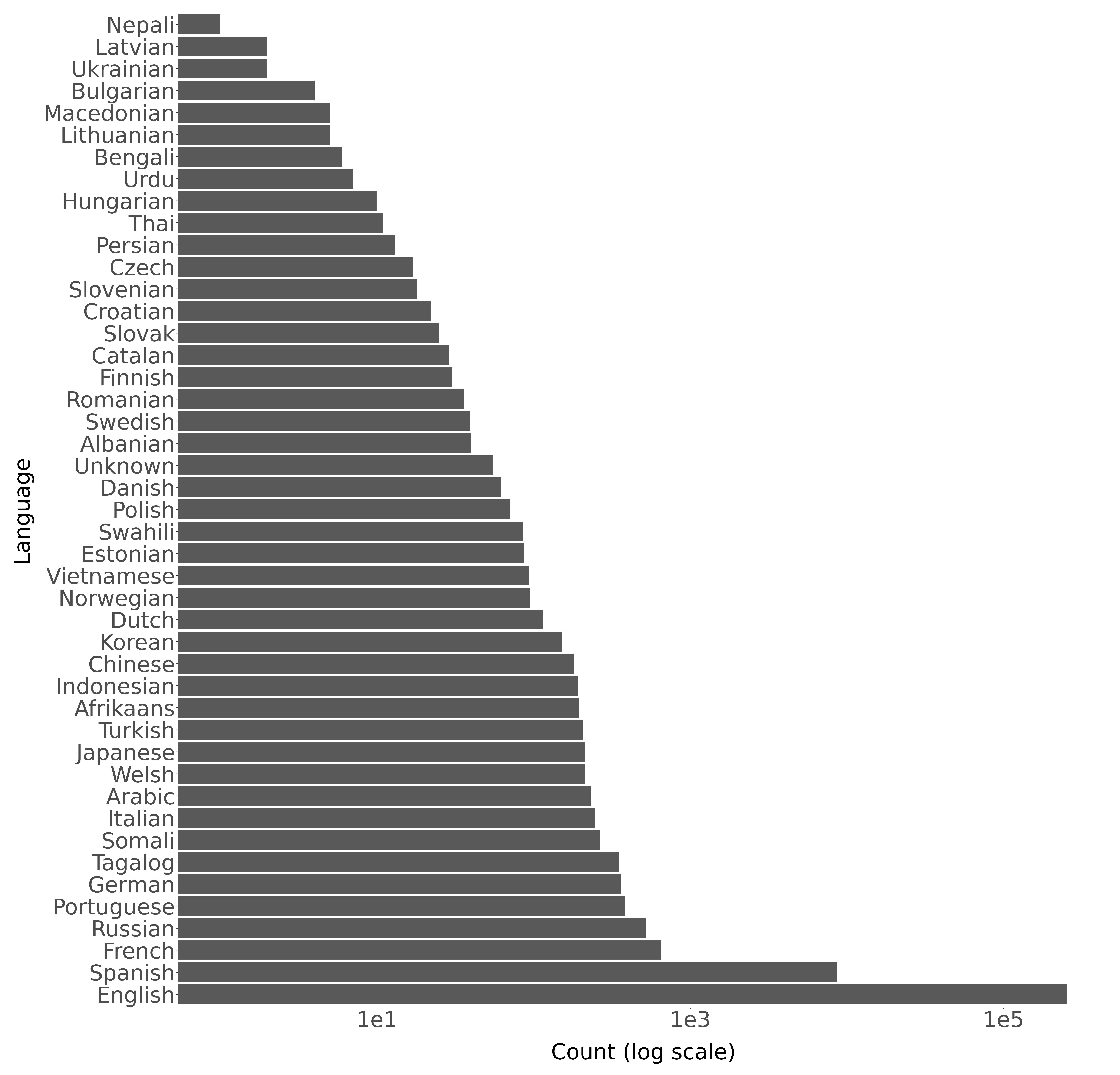}
  \caption{Classification of languages used within the transcripts (log scale)}
  \label{fig:langs}
\end{figure}

We applied LangDetect\footnote{\url{https://pypi.org/project/langdetect/}} on the transcripts for language detection. In Figure~\ref{fig:langs}, we show the languages within the transcripts generated by the TikTok Research API, excluding English. In our current dataset, we classified the 252,155 transcripts as English-based content. The second is Spanish-based, with 8,698 transcripts. In future work, we aim to provide an analysis of Spanish-based content to gain a more complete overview of the content published during the election cycle.

\subsection*{Most common bigrams in English and Spanish TikTok-generated transcripts.}
\begin{figure}[t]
  \centering
\includegraphics[height=0.35\textheight]{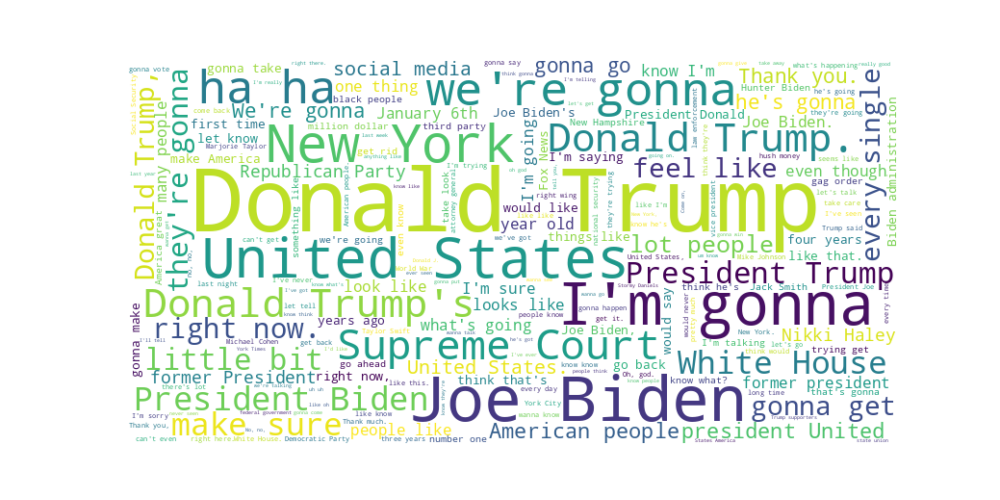}
  \caption{200 of the most frequent bigrams in the English Transcripts}
  \label{fig:eng_bigrams}
\end{figure}

\begin{figure}[t]
  \centering
\includegraphics[height=0.35\textheight]{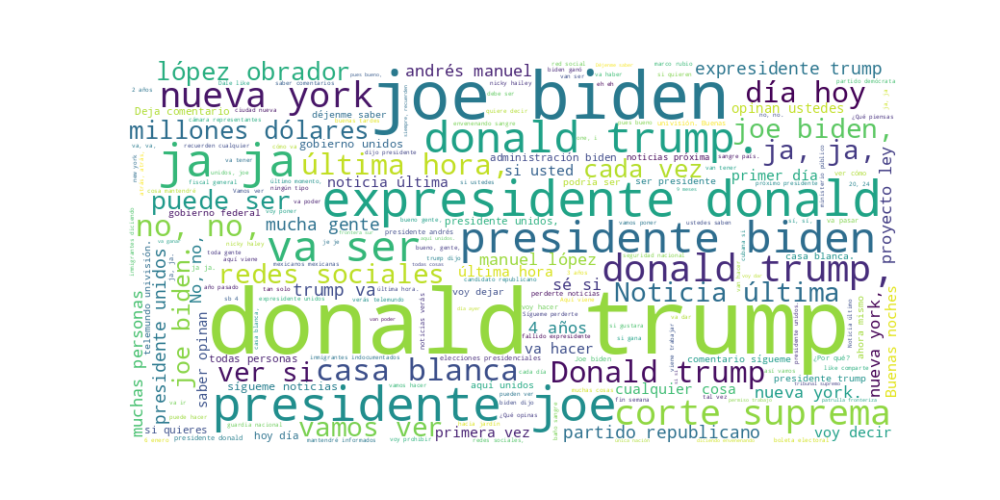}
  \caption{200 of the most frequent bigrams in the Spanish Transcripts}
  \label{fig:es_bigrams}
\end{figure}

We generated the bigrams for 252,155 of the English TikTok-generated transcripts and 8,698 of Spanish Tik-Tok generated transcripts, shown in Figure~\ref{fig:eng_bigrams} and Figure~\ref{fig:es_bigrams}, respectively.
Presidents Biden and Donald Trump are frequently mentioned based on the 200 most common bigrams in Spanish and English transcripts. Other common bigrams include Supreme Court ("corte suprema" in Spanish), New York ("nueva york" in Spanish), and social media ("redes sociales") in Spanish. In Spanish-based and English-based content, it is clear that the main topic discussed is related to the U.S. Presidential Elections. 


\section*{Conclusions}
This paper presents the current process of collecting TikTok content related to the upcoming U.S. Presidential Elections. The primary method for data collection used in this study was the TikTok Research API; however, due to the limitations mentioned earlier, we plan to use third-party scrapers for a thorough and complete analysis of the discourse. Based on our exploratory study on the 1,799,333 videos, the most frequent hashtags and keywords within the 'video\_description' attributes are related to President Joe Biden and Donald Trump. Within our dataset, 266,202 videos included a transcript generated by the TikTok API. Of these, 252,155 of the transcripts were detected to be written in English and 8,698 in Spanish. Within those transcripts, in addition to Donald Trump and President Joe Biden, the Supreme Court and New York were frequent bigrams mentioned in both transcripts.


\subsection*{Limitations}
The used API has certain limitations. The bearer token, used for authentication, remains valid for only two hours and must be regenerated after expiration to continue fetching the data. Each API call retrieves metadata for only 100 videos at a time. Additionally, when extracting more data for a specific date, a wait time must be included. If the search ID from a previous call is used to fetch subsequent records for the same day without this wait time, an error indicating an invalid search ID is thrown. Furthermore, a wait time must be observed after each request to comply with the API's rate limiting. Also, the API has a limit of 1,000 calls per day.

\subsection*{Future Work}
\begin{figure}[t]
  \centering
\includegraphics[width=0.8\textwidth]{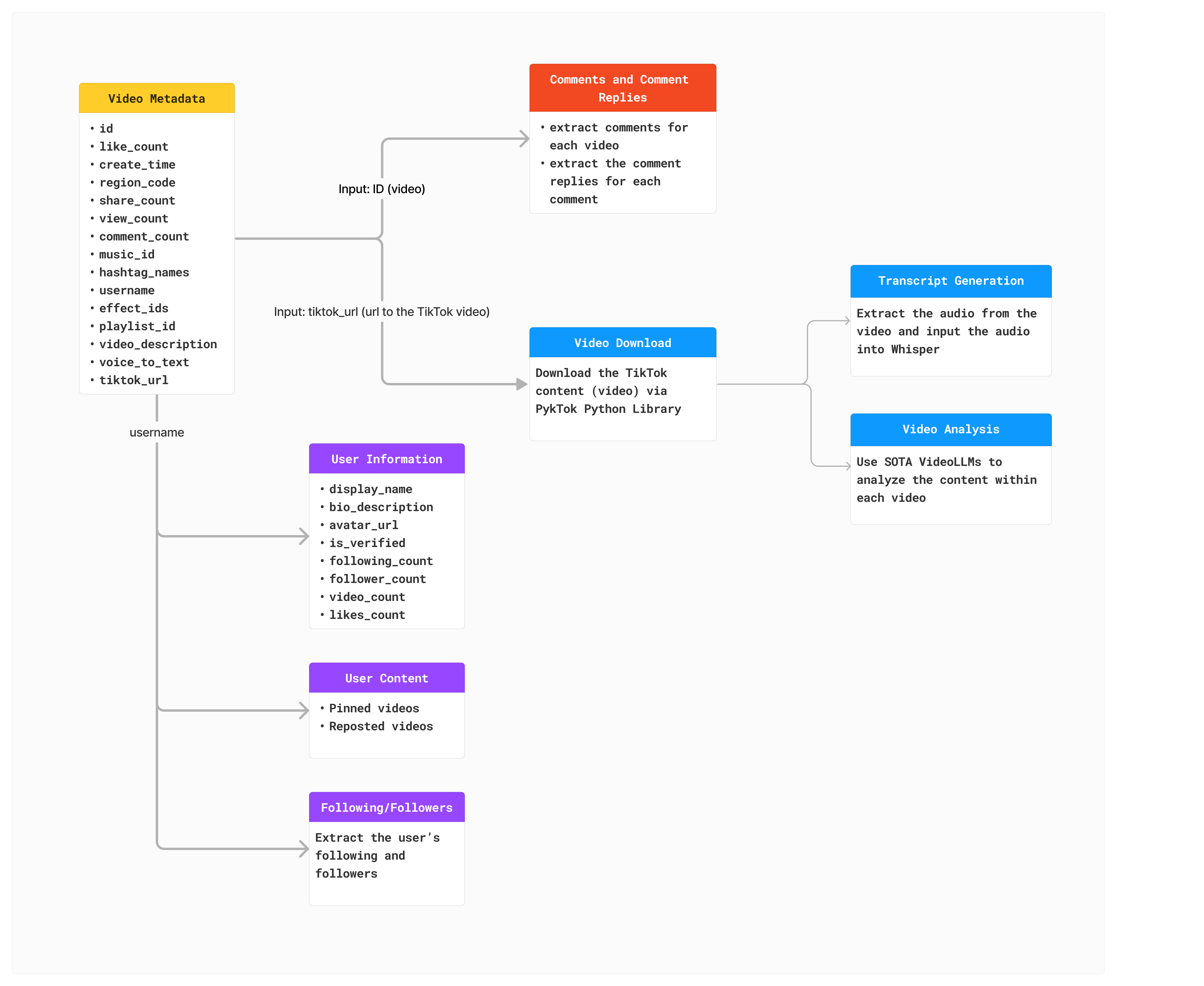}
  \caption{Data Collection Schema}
  \label{fig:dataprocess}
\end{figure}

To address the limitations in this paper and the gaps within our dataset, we will utilize third-party scrapers to collect more video metadata. To conduct a more in-depth analysis, we also plan to collect other attributes within a video, as shown in Figure~\ref{fig:dataprocess}. While collecting more metadata, we are in the process of collecting videos to analyze the content using VideoLLMs such as Video-LLaMa\cite{damonlpsg2023videollama} or LLaVa-NeXT-Video\cite{liu2024llavanext}. We will collect the comments and comment replies for each video using a third-party scraper to study the discussions between users within the comment section. Due to the low presence of transcripts for each video, we will extract the audio and use Whisper to generate the transcript for each video.


\section*{Data Availability} 

Our data collection will continue uninterrupted for the foreseeable future. 
As the election approaches, we anticipate that the amount of data will grow significantly. The data set available on GitHub is released in compliance with the Tiktok's Terms and Conditions, under which we are unable to publicly release the videos of the collected posts. We are, therefore, releasing the Video IDs, which are unique identifiers tied to specific posts. The Video IDs can be used by researchers to query Tiktok’s API and obtain the complete video objects, including multimedia and metadata information as depicted in Figure \ref{fig:dataprocess}.
A publicly accessible GitHub repository that we will continue to routinely update is available at \href{https://github.com/gabbypinto/US2024PresElectionTikToks}{https://github.com/gabbypinto/US2024PresElectionTikToks}

\section*{About the Team}
The 2024 Election Integrity Initiative is carried out by a collective of USC students and volunteers whose contributions are instrumental to enable these studies. The authors are grateful to the following HUMANS Lab's members for their tireless efforts on this project: Ashwin Balasubramanian, Leonardo Blas, Keith Burghardt, Sneha Chawan, Vishal Reddy Chintham, Eun Cheol Choi, Srilatha Dama, Priyanka Dey, Isabel Epistelomogi, Saborni Kundu, Grace Li, Richard Peng, Jinhu Qi, Ameen Qureshi, Namratha Sairam, Srivarshan Selvaraj, Kashish Atit Shah, Gokulraj Varatharajan, Reuben Varghese,  Siyi Zhou, and Vito Zou.

\bibliographystyle{ACM-Reference-Format}
\bibliography{chatgpt, genai,tiktok}

\newpage
\appendix
\section*{Appendix}

\paragraph{Table \ref{tab:words_hashtags}: Keywords and hashtags applied in the query with respect to its publication date.}

\begin{tiny}
\begin{longtable}{|c|c|c|c|c|c|}
\hline
\textbf{Keywords/Hashtags} & \textbf{Phase 1} & \textbf{Phase 2} & \textbf{Phase 3} & \textbf{Phase 4} & \textbf{Phase 5} \\ 
\hline
\endfirsthead
\multicolumn{6}{c}{{\tablename\ \thetable{} -- continued from previous page}} \\
\hline
\textbf{Keywords/Hashtags} & \textbf{Phase 1} & \textbf{Phase 2} & \textbf{Phase 3} & \textbf{Phase 4} & \textbf{Phase 5} \\ 
\hline
\endhead
\hline \multicolumn{6}{|r|}{{Continued on next page}} \\ \hline
\endfoot
\hline
\endlastfoot
election2024 & - & - & - & + & + \\ 
\hline
Election2024 & - & - & - & + & + \\ 
\hline
US Elections & + & + & + & + & + \\ 
\hline
USElections & + & + & + & + & + \\ 
\hline
us elections & + & + & + & + & + \\ 
\hline
uselections & + & + & + & + & + \\ 
\hline
2024Elections & + & + & + & + & + \\ 
\hline
2024 Elections & + & + & + & + & + \\ 
\hline
2024 elections & + & + & + & + & + \\ 
\hline
2024elections & + & + & + & + & + \\ 
\hline
2024election & - & + & + & + & + \\ 
\hline
2024PresidentialElections & + & + & + & + & + \\ 
\hline
2024 Presidential Elections & + & + & + & + & + \\ 
\hline
2024presidentialelections & + & + & + & + & + \\ 
\hline
2024 presidential elections & + & + & + & + & + \\ 
\hline
saveamerica2024 & - & - & + & + & + \\ 
\hline
presidentbiden & - & - & + & + & + \\ 
\hline
Biden & + & + & + & + & + \\ 
\hline
biden & + & + & + & + & + \\ 
\hline
bidenharris & - & + & + & + & + \\ 
\hline
JoeBiden & + & + & + & + & + \\ 
\hline
Joe Biden & + & + & + & + & + \\ 
\hline
joebiden & + & + & + & + & + \\ 
\hline
joe biden & + & + & + & + & + \\ 
\hline
joseph biden & + & + & + & + & + \\ 
\hline
Joseph Biden & + & + & + & + & + \\ 
\hline
Biden2024 & + & + & + & + & + \\ 
\hline
biden2024 & + & + & + & + & + \\ 
\hline
bidenharris2024 & - & - & + & + & + \\ 
\hline
Donald Trump & + & + & + & + & + \\ 
\hline
donald trump & + & + & + & + & + \\ 
\hline
DonaldTrump & + & + & + & + & + \\ 
\hline
donaldtrump & + & + & + & + & + \\ 
\hline
donaldtrump2024 & - & + & + & + & + \\ 
\hline
Trump2024 & + & + & + & + & + \\ 
\hline
trump2024 & + & + & + & + & + \\ 
\hline
trumpsupporters & + & + & + & + & + \\ 
\hline
trumptrain & + & + & + & + & + \\ 
\hline
republicansoftiktok & + & + & + & + & + \\ 
\hline
conservative & + & + & + & + & + \\ 
\hline
MAGA & + & + & + & + & + \\ 
\hline
maga & + & + & + & - & - \\ 
\hline
makeamericagreatagain & - & + & + & + & + \\ 
\hline
ultramaga & - & + & + & + & + \\ 
\hline
KAG & + & + & + & + & + \\ 
\hline
Republican & + & + & + & + & + \\ 
\hline
trump2024 & - & + & + & + & + \\ 
\hline
presidenttrump & - & - & + & + & + \\ 
\hline
trumpismypresident & - & - & + & + & + \\ 
\hline
letsgobrandon & - & + & + & + & + \\ 
\hline
GOP & + & + & + & + & + \\ 
\hline
CPAC & + & + & + & + & + \\ 
\hline
NikkiHaley & + & + & + & + & - \\ 
\hline
nikkihaley & + & + & + & + & - \\ 
\hline
DeSantis & + & - & - & - & - \\ 
\hline
RonDeSantis & + & - & - & - & - \\ 
\hline
desantis & + & - & - & - & - \\ 
\hline
rondesantis & + & - & - & - & - \\ 
\hline
RNC & + & + & + & + & + \\ 
\hline
democratsoftiktok & + & + & + & + & + \\ 
\hline
democratsarehot & + & + & + & + & + \\ 
\hline
thedemocrats & + & + & + & + & + \\ 
\hline
voteblue2024\blueheart & - & - & + & + & + \\ 
\hline
voteblue2024 & - & - & + & + & + \\ 
\hline
vote blue & - & - & + & + & + \\ 
\hline
DNC & + & + & + & + & + \\ 
\hline
dnc & + & + & + & + & + \\ 
\hline
kamalaharris & + & + & + & + & + \\ 
\hline
kamala harris & - & - & - & + & + \\ 
\hline
mariannewilliamson & + & + & - & - & - \\ 
\hline
deanphillips & + & + & + & - & - \\ 
\hline
williamson2024 & + & + & + & - & - \\ 
\hline
phillips2024 & + & + & + & - & - \\ 
\hline
democratic party & + & + & + & + & + \\ 
\hline
Democratic party & + & + & + & + & + \\ 
\hline
republican party & + & + & + & + & + \\ 
\hline
Republican party & + & + & + & + & + \\ 
\hline
Third party & + & + & + & + & + \\ 
\hline
third party & + & + & + & + & + \\ 
\hline
Green party & + & + & + & + & + \\ 
\hline
green party & + & + & + & + & + \\ 
\hline
Independent party & + & + & + & + & + \\ 
\hline
independent party & + & + & + & + & + \\ 
\hline
No Labels & + & + & + & + & - \\ 
\hline
RFKJr & + & + & + & + & + \\ 
\hline
RFK Jr. & + & + & + & + & + \\ 
\hline
RFK Jr & + & + & + & + & + \\ 
\hline
rfkjr & + & + & + & + & + \\ 
\hline
rfkjr. & + & + & + & + & + \\ 
\hline
rfk & + & + & + & + & + \\ 
\hline
Robert F. Kennedy Jr. & + & + & + & + & + \\ 
\hline
robert f. kennedy Jr. & + & + & + & + & + \\ 
\hline
jill stein & + & + & + & + & + \\ 
\hline
jillstein & + & + & + & + & + \\ 
\hline
Jill Stein & + & + & + & + & + \\ 
\hline
JillStein & + & + & + & + & + \\ 
\hline
CornellWest & + & + & + & + & + \\ 
\hline
cornellwest & + & + & + & + & + 
\label{tab:words_hashtags}
\end{longtable}
\end{tiny}

\end{document}